\RequirePackage{fixltx2e}
\documentclass[%
 reprint,
 amsmath,amssymb,aps, showpacs, 
]{revtex4-1}
\usepackage[english]{babel}
\usepackage{enumitem}
\usepackage{fancyhdr}
\usepackage{amsfonts, amssymb}
\usepackage{graphicx}
\usepackage{amsmath}   
\usepackage{graphicx}
 \usepackage{url}
  \usepackage{tikz}
 \usepackage{hyperref}
\usepackage{dcolumn}
\usepackage{placeins}

\newcommand{\dist}{\textrm{dist}}

\renewcommand{\c}[1]{\mathcal{#1}}
\renewcommand{\t }[1]{\mathrm{#1}}

\tikzstyle{no1}=[circle, draw, fill = yellow!50!green, inner sep = .08 cm]
\tikzstyle{aux} =[ inner sep = .08 cm]
\tikzstyle{no2}=[rectangle, draw, fill = yellow!0!green, inner sep= .08 cm]
\tikzstyle{no3}=[circle, draw, fill = yellow!95!green, minimum size= .35 in]
\tikzstyle{no4}=[circle, draw, fill = yellow, inner sep= .08 cm ]

\tikzstyle{circle2}=[dashed, line width = 1]
\tikzstyle{circle3}=[line width = 6.9, color = {rgb:black,1;white,1}, opacity = .5]
\tikzstyle{no1}=[circle, draw, fill = yellow!50!green, inner sep = .08 cm]
\tikzstyle{aux} =[ inner sep = .08 cm]
\tikzstyle{no2}=[rectangle, draw, fill = yellow!0!green, inner sep= .08 cm]
\tikzstyle{no3}=[circle, draw, fill = yellow!95!green, minimum size= .35 in]
\tikzstyle{no4}=[circle, draw, fill = yellow, inner sep= .08 cm ]

\begin{document}
\title{Growth and Containment of a Hierarchical Criminal Network}
\author{Charles Z. Marshak}
\affiliation{UCLA Department of Mathematics}

 \author{M. Puck Rombach}
 \author{Andrea L. Bertozzi}
 \affiliation{UCLA Department of Mathematics and Institute of Pure and Applied Mathematics}
  \author{Maria R. D'Orsogna}
  \email{dorsogna@csun.edu}
   \affiliation{UCLA Department of Biomathematics and CSUN Department of Mathematics}
  \
\date{\today}

\begin{abstract}
\noindent
We model the hierarchical evolution of an organized criminal network
via antagonistic \textit{recruitment} and \textit{pursuit} processes.
Within the recruitment phase, a criminal kingpin enlists new members
into the network, who in turn seek out other affiliates. New recruits
are linked to established criminals according to a probability
distribution that depends on the current network structure. At the
same time, law enforcement agents attempt to dismantle the growing
organization using pursuit strategies that initiate on the lower level
nodes and that unfold as self-avoiding random walks.  The global
details of the organization are unknown to law enforcement, who must
explore the hierarchy node by node. We halt the pursuit when certain
local criteria of the network are uncovered, encoding if and when an
arrest is made; the criminal network is assumed to be eradicated if
the kingpin is arrested.  We first analyze recruitment and study the
large scale properties of the growing network; later we add pursuit
and use numerical simulations to study the eradication probability in
the case of three pursuit strategies, the time to first eradication
and related costs. Within the context of this model, we find that
eradication becomes increasingly costly as the network increases in
size and that the optimal way of arresting the kingpin is to intervene
at the early stages of network formation. We discuss our results in
the context of dark network disruption and their implications on
possible law enforcement strategies.

\end{abstract}
\pacs{89.65.Ef, 89.75.Hc, 87.23.Ge}

\maketitle

\section{Background}

\begin{figure*}
\centering
The Recruitment Process
\includegraphics[width = 1\linewidth]{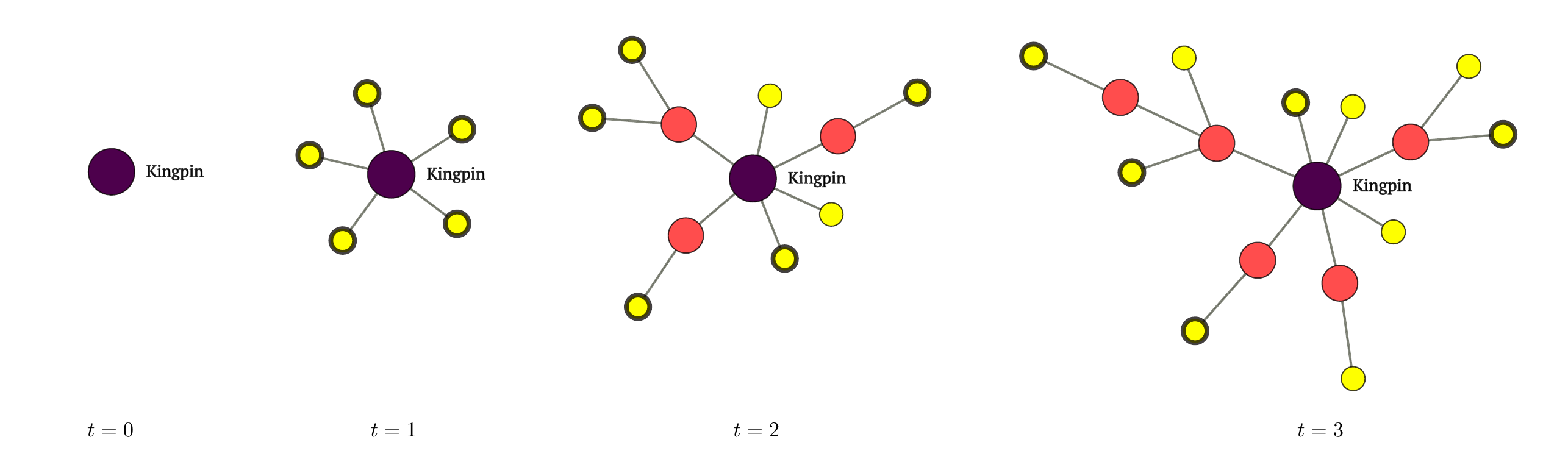}
\caption{A simulated recruitment process for $t = 0,1, 2, 3$.  The
  network starts with a single criminal, the kingpin, and then evolves
  according to the preferential attachment mechanism described in the
  text.  Here the number of new criminals introduced into the network
  is given by the recruitment index $k=5$.  The lighter yellow colored nodes
  represent criminals without underlings, and will be referred to as
  street criminals.  Of these, the nodes with the darker boundary are
  those freshly recruited at a given time step.  For example, the
  number of street criminals when $t = 3$ is 8, of which five are new
  recruits.}
\label{d3evo}
\end{figure*}

\noindent
Modeling criminal socio-economic systems using tools from statistical
mechanics, complex networks, partial differential equations and game
theory, and building on well established social and behavioral
phenomena has become of great interest in recent years
\cite{econsocbook, castellano, review}.  This interdisciplinary effort
has helped shed light on the formation of crime hotspots \cite{short1,
  bert2, hotspots}, the dynamics of criminal behavior \cite{maria2012,
  bellomo2014}, the mechanics of gang rivalries \cite{alexey} and
recidivism trends \cite{recidivism}.  In this paper, we bring some of these tools to
the problem of organized crime.

Underworld syndicates can be quite
successful in profiting from exploitation, theft, intimidation, and
murder through complex webs of disciplined units governed
via authority structures, division of labor, and strict
behavioral codes.  Conversely, governments must effectively protect
their citizens, maintain internal order, and safeguard the channels of
the legal economy. Since organized criminal associations mostly operate in sophisticated
and secretive ways, a pertinent way of studying them is through
the well-known framework of dark networks \cite{mcbride3, duijn}.
Here, actors and links are predominantly hidden to possible
disruptors.  Criminals must decide who to interact with and how to
balance the threat of possibly being arrested with the profits
afforded by criminal collaboration \cite{mcbride3}.  On the other
hand, police agents must eradicate the criminal enterprise often without an a priori
full knowledge of the global network. 

Within the context of criminal behavior, the disruption of dark networks has been studied via
probability distributions that embody the limited information
available to law enforcement agents and that govern dismantling
attempts \cite{mcbride1, mcbride2, mcbride3}.  Other models analyze network
eradication when the entire structure is known \cite{ball,
  borgattikey}. Typically, each criminal is assigned a utility value
based on his or her prominence in the organization, which is then used
by law enforcement to orchestrate optimal intervention strategies.
Several disruption tactics were tested on a well-documented drug
trafficking network in the Netherlands \cite{duijn}. Here, disruption
was modeled as a process of node removal from a dynamic network, where
some of the disruption strategies assume full knowledge of the
network. New players are not introduced, but once a node has been
eliminated, the network is allowed to ``recover" and links among nodes
may readjust. The authors found that interventions are most effective
at the very early stages of the disruptive process, since, in the long
run, perturbations and reorganizations lead to a more robust and
resilient network. In other work, data from well-known transnational terrorist and
criminal networks were used to simulate different disruption
strategies, comparing the removal of ``bridge" and ``hub" nodes
\cite{dark_topo}.  All models described above offer insight into fully visible or dark
network disruption, but none of them factor in growth or
recruitment.

The goal of this paper is to fill this void and to study criminal
networks as they expand, in an effort to increase reach and
profitability, while law enforcement engages in their suppression.
Different types of networks can be studied, such as white-collar,
racketeering or terrorist organizations; for concreteness we focus on
vertically organized crime networks, such as the Central and South
American drug cartels whose constituents are usually structured by
ranks of influence, seniority and activity \cite{cartels_, mexico,
  sinaloa1, sinaloa2}.
We thus study disruption strategies on hierarchical, \emph{growing},
dark criminal networks. Within the context of this work, a hierarchal criminal network is
defined to be one where every criminal has precisely one link to a
more senior member, except for the organization's head, whom we refer
to as the kingpin. The network links are thus assumed to be professional
connections, not social ties. 
Concurrent to law enforcement eradication attempts, we
include a mechanism for criminal recruitment. 
As we shall see, the hierarchal 
structure and the interplay between the two antagonistic trends --
recruitment and disruption -- lead to interesting dynamics and
implications on the optimal strategies to be used in
eradicating the criminal organization.

To model recruitment we use a variation of preferential
attachment models that were originally proposed to study the topology of the internet \cite{BA1, BA2}
and that are now implemented in a variety of contexts
\cite{BA1, BA2, socialdist, pasex, tonci, Mahmoud1, Mahmoud2,
  mason2014}.  The central assumption,
roughly speaking, is that the ``rich get richer", meaning that
webpages with many existing links are more likely to be connected to
newly introduced ones.  We use these preferential attachment models to guide
the design and analysis of our recruitment mechanism. Particularly useful is the
notion of a ``social distance"   \cite{socialdist, socialdist2} that we will use to model
criminal recruitment. Here, each node of a static network is associated to a set of socially
relevant features such as profession, religion or location that can be used to construct
an ad-hoc metric quantifying the ``social" distance between nodes, 
and that can be different from the topological distance. 

The aim of law enforcement is to disrupt and
possibly eradicate the criminal network by capturing the kingpin, as
successfully accomplished by the Colombian government in collaboration
with the Drug Enforcement Administration (DEA) in the case of the
Medell\'{i}n cartel \cite{crime1}. According to the DEA, a major
factor leading to the collapse of Pablo Escobar's drug organization
was the so called ``kingpin strategy", where the senior cartel members
overseeing network operations were specifically targeted. The Mexican
government is pursuing the same strategy in its current war on drugs
against the Sinaloa, Gulf, Juarez, Tijuana, Beltran Leyva and
Guerreros Unidos cartels \cite{kingpin_mex}, although with mixed
results.
The arrest mechanism we utilize draws on probabilistic node and link removal processes 
including cascades that arise when the elimination of a single node triggers the removal of others.  Node removal cascades are often used to model wireless networks and power grids where the failure of a tower may isolate others \cite{cascades1, cascades2}; they have also been adapted and used to model contagion \cite{dis1}, neuronal \cite{av0,av1, av2}, and terrorist networks \cite{gutfraind5}.  In this work, we will taylor node removal processes to represent criminal arrest by law enforcement. 

While all the applications described above are not necessarily related to organized crime, we will draw upon these many different perspectives to best model criminal recruitment and dark network disruption on the model we describe below.

\section{Overview}

\noindent
In this section, we present an overview of our dynamical, hierarchal criminal network stemming from a so-called kingpin.  Our model alternates between two key processes: the recruitment of criminals to the network and the concurrent, antagonistic pursuit and disruption by law enforcement. 

To recruit new criminals into the network, we use a preferential attachment mechanism, a schematic of which is shown in Fig.\,\ref{d3evo}.  Criminal nodes are added to the network at a constant rate, each forming a link to a more senior member.  Here, we assume that there is a large pool of potential new members the criminal network can recruit from, as documented for Central and South American drug cartels \cite{williams, cartels_2, mexico, oc_book, sinaloa1, sinaloa2}.  We assume the entire network structure to be initially hidden to law enforcement except for the visible ``street-level" criminals at the end of the network that do not have any further underlings. These street criminals are the ones that are the most directly involved in drug dealings while more nested members of the hierarchy are assumed to act more like masterminds: the higher up a criminal is in the network, the less likely he or she is to overtly engage in criminal enterprises, effectively shielding themselves from criminal implications \cite{cartels_, oc_book}.  We thus assume that police intervention must begin from street criminals,  later progressing to higher nodes, so that the network structure becomes visible to law enforcement gradually, in a node by node fashion.  

The distance between a given criminal node on the network and visible street-level activity is defined as the smallest number of connections separating the given node from any street criminal. The closer a criminal is to visible street-level activity, the more vulnerable he or she is to detection and arrest.  As such, senior criminals will seek to maintain a buffer between themselves and street criminals  \cite{theft_of_nation, kenney, morselli_cont}. Due to their visibility, the latter are assumed to have greater access to potential recruits and to aggressively seek new underlings in order to avoid their own exposure to law enforcement.  Combining these two heuristics, we posit that within the recruitment process,  prospective criminals are most likely to establish a link with street criminals, rather than with more nested members of the criminal hierarchy. This description of recruitment is echoed in Refs.\, \cite{williams, recruitment1, mafia_nyc, oc_book, cartels_}, and a precise mathematical formulation will be given in the next section.  

As mentioned above, law enforcement agents can pursue the kingpin starting only from street criminals, progressively moving to more nested, connected links. We let each node  encountered by law enforcement to be subject to an ``investigation," and impose that the overall motion of an agent be represented by a self-avoiding random walk. 
Since agents have access to incomplete or inaccurate information, their movements  may appear random to an observer with perfect information of the organization's layout. 
At any point during the pursuit, the criminal under investigation can be ``arrested"  with all of its associated underlings. Although we refer to node removal as arrest, the latter may represent exile, extradition, or assassination \cite{pablo1, pablo2}.  By removing criminals from the network, the kingpin becomes more vulnerable to future capture. However, since the overall structure of the network is unknown to law enforcement,  the random walk may also lead to a dead-end: in moving from node to node, the officer may reach a new street criminal without underlings, with no further investigation possible.  In this case, the pursuit is terminated and deemed unsuccessful.  If instead the kingpin is reached and arrested,  the criminal network is assumed to be eradicated.

While our modeling assumptions are necessarily simplified for mathematical analysis, they are motivated from in-the-field methodologies and outcomes used by governmental agencies. For example, when the DEA employs its ``buy and bust" strategy to bait high level drug traffickers \cite{wilson}, certain metrics are used to assess the value of a possible arrest, including the criminal's influence and the likelihood of reaching the kingpin \cite{oc_book}.  On certain occasions these sequences of investigations lead to the arrest of high ranking drug traffickers \cite{DEA_good}, while in other circumstances they yield dead-ends, resulting in public embarrassment \cite{DEA_bad}.  

Finally, before illustrating our recruitment and arrest model more in detail, in Table \ref{table1}  we list standard network terminology and the corresponding nomenclature used here, for context. For example, in standard network terminology, our network is a tree,  the street criminals are the leaves and the kingpin is the root.

\section{Recruitment}

\begin{table}\centering
\begin{tabular}{|p{4cm}|p{4cm}|}
\hline
This model &Network theory\\
\hline
\hline
Kingpin& Root\\
\hline
Underlings of criminal $j$ & Children of node $j$\\
\hline
Criminal network & Rooted directed tree \\
\hline
$\c C{(t)}$: Criminals (including kingpin) at time $t$ & Vertices or nodes (including root) at time $t$\\
\hline
$\c S{(t)}$: Set of criminals without underlings  & Set of leaves at time $t$, i.e. nodes of out degree $0$ \\
\hline
\end{tabular}
\caption{A table comparing the terminology used in this paper and that of standard network theory.}
\label{table1}
\end{table}

\noindent
We now focus on the mathematical aspects of the recruitment process. 
We start with an initial criminal network formed solely by the kingpin. The network evolves recursively so that at time $t$ it contains a set $\c C{(t)}$ of criminals, including the kingpin.  Of these, the subset $\c S{(t)}$ denotes street criminals, those without any underlings.  We also introduce the metric dist$(j;t)$ to denote the distance separating criminal $j$ from street activity  and defined as the minimum number of links between criminal $j$ and any other street criminal in the hierarchy beneath it.   At every time step increment, from $t$ to $t+1$, we add $k$ new recruits to the network according to a preferential attachment mechanism.  Every node $j$ is assigned a weight $w(j;t)$ to quantify the relative likelihood that it will link a new criminal underling. Since, as discussed above, a plausible assumption is that street criminals are the most likely to recruit new criminals into the organization, we let $w(j;t)$ be inversely proportional to the distance between $j$ and ${\cal S}(t)$ so that

\begin{align}
w(j;t)  &= \frac{1}{\dist(j;t)+a},
\end{align}

\noindent
where $a$ is a parameter, which we set to $a=1$ for simplicity.  With these choices, $w(j;t)$ embodies the proximity of criminal $j$ to visible street-level activity on a scale from zero to one, with one being the closest possible.  If criminal $j$ is a street criminal, then dist$(j;t)=0$ and $w(j;t) = 1/a=1$, the maximum possible value. On the other hand, as the network keeps growing, higher level nodes can become progressively detached from street activity, so that in principle dist$(j;t) \to \infty$ and $w(j;t) \to 0$. Note that the choice $a \to \infty$ leads to a uniform weight $w(j;t)$ for all criminals. In this case, the recruitment process can be described as the growth of a recursive tree \cite{Mahmoud1, Mahmoud2}.  On the other extreme, the choice $\displaystyle{a \to 0}$ would restrict the recruitment process exclusively to street criminals, barring higher ranking criminals from adding subordinates.  Our decision to use a finite, non zero value $a  = 1$ ensures that recruitment is not exclusive to $\c S(t)$.   

\begin{figure}
{The Criminal Network and $w(j;t)$}
\includegraphics[width =.85\linewidth]{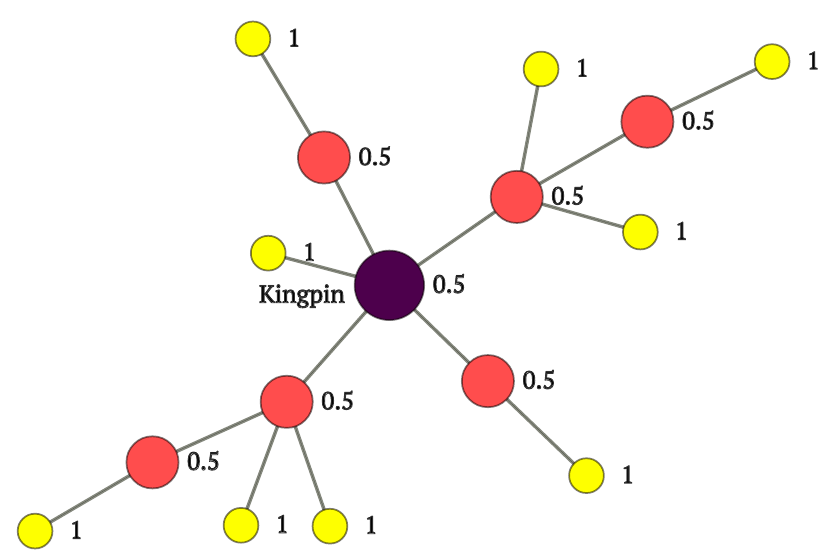}
\caption{The criminal network at $t = 3$ with the values of $w(j;t)$ explicitly shown. Here, the recruitment index $k=5$ and the initial configuration was that of a single kingpin.  All criminals $j$ within $\c S(t)$ have weight $w(j;t)= 1$ since on these nodes $\dist(j;t) = 0$.}
\label{weights}
\end{figure}

After evaluating $w(j;t)$ for all existing nodes, we iteratively introduce $k$ new criminals to the network. We add them one by one to nodes
that are selected according to the relative weights $w(j;t)$. Note that each existing criminal can add multiple underlings within a single time step, since the recruitment of one new member does not exclude the possibility of a different new member being recruited by same criminal. 
We call $k$ the recruitment index. 
In Fig.\,\ref{weights} we depict a particular network configuration, including the explicit weights $w(j;t)$ assigned to each criminal $j$.

\begin{table}\centering
  \begin{tabular}{|m{2cm}|p{5cm}|}
\hline
Parameter&Description\\
\hline
\hline
$t$& Time\\
\hline
$k$ & Recruitment index -- number of new criminals added to the network at each time step \\ 
\hline
$a$& Node weight parameter in $w(j;t)$.\\
\hline
$n$& Number of criminals on network at time $t$, $n = k t +1$ \\
\hline
\end{tabular}
  \caption{The parameters of the recruitment mechanism.}
  \label{parameters}
\end{table}

\subsection{Out Degree Distribution}

\begin{figure}
\centering
\includegraphics[width=\linewidth]{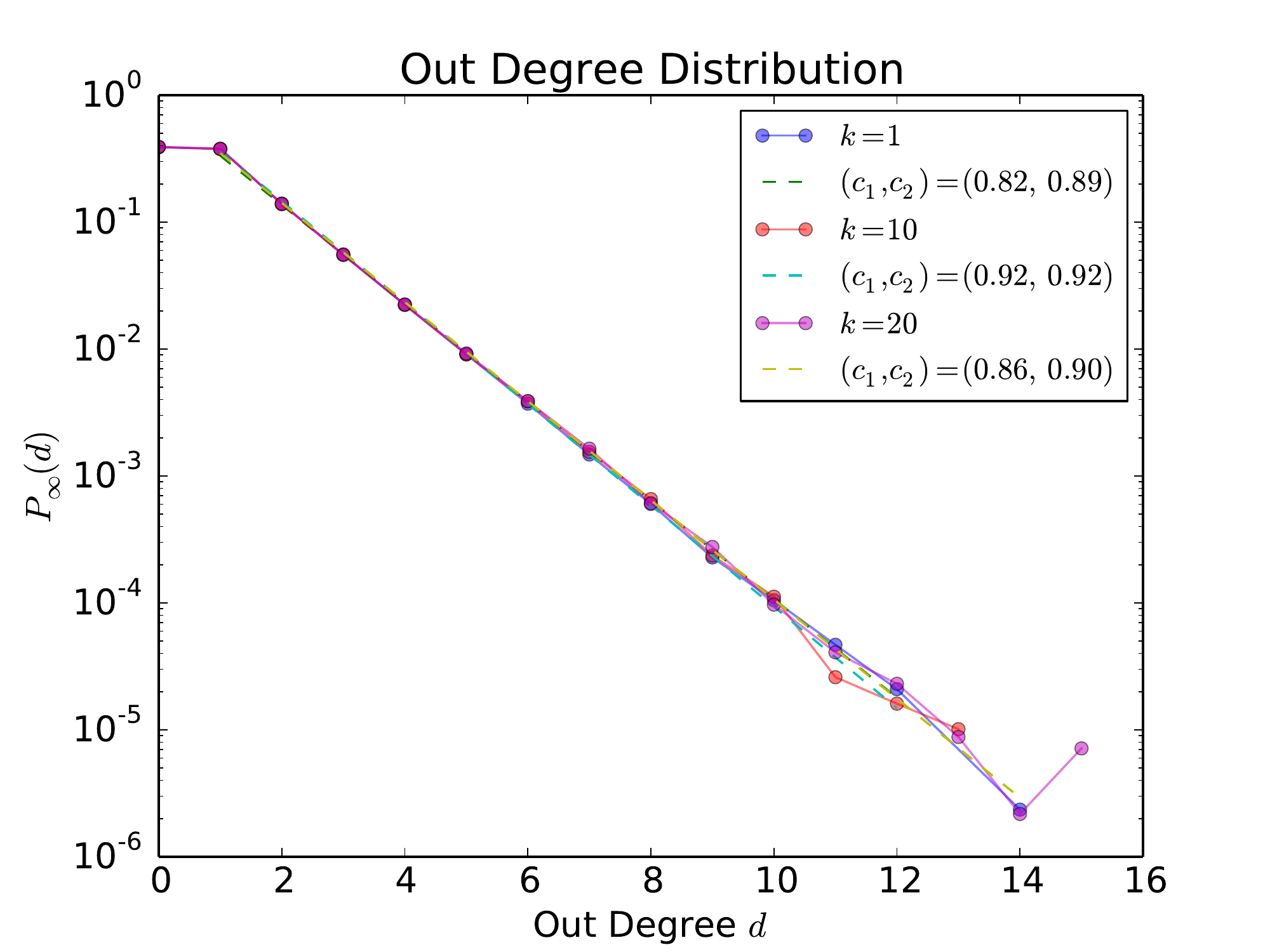}
\caption{The out degree distribution $P_{\infty} (d)$ of nodes on a criminal network determined from numerical simulations for $t \to \infty$. The three curves correspond to 
recruitment indices $k = 1, 10, 20$. Simulations were terminated when the total number of criminals exceeded $5 \times10^3$. All curves for
$P_{\infty} (d)$ are averaged over 100 runs.  The tail of the degree distribution is noisy due high degree nodes occurring sporadically. Running the simulations for longer times will extend the range of the domain in $d$, but not the form of $P_{\infty} (d)$.  We conjecture the degree distribution to follow an exponential law independent of $k$, and fit it to a decaying exponential distribution as discussed in the text. The value for $P_{\infty}(d=0)$ = 0.39 for all values of $k$.}
\label{DD}
\end{figure}

\noindent
We can now investigate the statistics related to our recruitment model.  First, we explore how the total number of underlings a criminal has varies throughout the network, \textit{i.e.} we analyze the \emph{out degree probability distribution}. By out degree we indicate the number of nodes in the hierarchy directly beneath a criminal, excluding higher nodes from the enumeration.  For example, in Fig.\,\ref{weights} the out degree of the kingpin is 5 criminals. Of these five, the upper left two have out degree one and the other three have out degree two. 
We do not impose any limitations on the number of underlings connected to any given node, either directly or indirectly. In principle thus, a node can have an infinitely large number of subordinates.  However, due to the choices made in modeling the attachment weights $w(j;t)$ we expect that as the organization grows and more senior criminals become more entrenched within the network, their likelihood of adding new recruits decreases in favor of criminals that are closer to street activity. As a result, we expect our organized crime network to grow several hierarchal levels and to have a few key players, such as the kingpin, who directly oversee a relatively large number of criminals while the rest have at most one or two underlings that ``report" to them. 
We introduce the time dependent out degree probability distribution $P(d;t)$ for a randomly selected node to have $d$ direct underlings at time $t$.
At the onset of network growth, when the only criminal present is the kingpin, $P(0;0) = 1$.  As the number of added nodes $n = kt +1 $ increases, 
we conjecture that for large enough $d$, $P(d;t \to \infty)$ can be approximated via an exponential form

\begin{eqnarray}
\label{distribute}
P_{\infty}(d) \equiv P(d;t \to \infty) =   c_1e^{- c_2 d}, 
\label{ddd} 
\end{eqnarray}

\noindent
for constants $c_1, c_2$. Following the above discussion on the nature of $w(j;t)$, we expect that as $t \to \infty$ most of the new underlings will be connected to existing street criminals, resulting in $P_{\infty}(d=0) \simeq P_{\infty}(d=1)$.  Using this approximation and Eq.\,\ref{distribute} for $d  \geq 1$, we expect $c_1, c_2, P_\infty(d=0)$ to be related by

\begin{eqnarray}
\label{match}
c_1 \simeq [1 - P_{\infty}(d=0)](e^{c_2} - 1).
\end{eqnarray}

\noindent
In Fig.\,$\ref{DD}$, we grew the network to 5 $\times$ $10^3$ criminals and found Eq.\,\ref{distribute} to accurately describe the out degree distribution for $d \geq 1$,
with Eq.\,\ref{match} being accurate to first approximation. 
We also varied $k$ between 1 and 20 and did not notice substantial variations in fitted parameter values.  Moreover, we considered smaller sized networks (not shown here) with 500 criminals and found that Eq.\,\ref{distribute} still described the data well. These results suggest that at large times, network structure is independent of its rate of growth. In particular, as $t \to \infty$ our results show that the probability that any given node has no underlings is given by the $k$-independent, universal value $P_{\infty}(d=0) = 0.39$. The average out degree $d$ is given by

\begin{eqnarray}
\label{avg}
\langle d \rangle  = c_1 \sum_{d=1}^{\infty} d e^{-c_2 d} = \frac{1 - P_{\infty}(d=0)}{1-e^{-c_2}}.
\end{eqnarray}

\subsection{Criminal Density and Position}

\begin{figure}
\centering
\includegraphics[width=\linewidth]{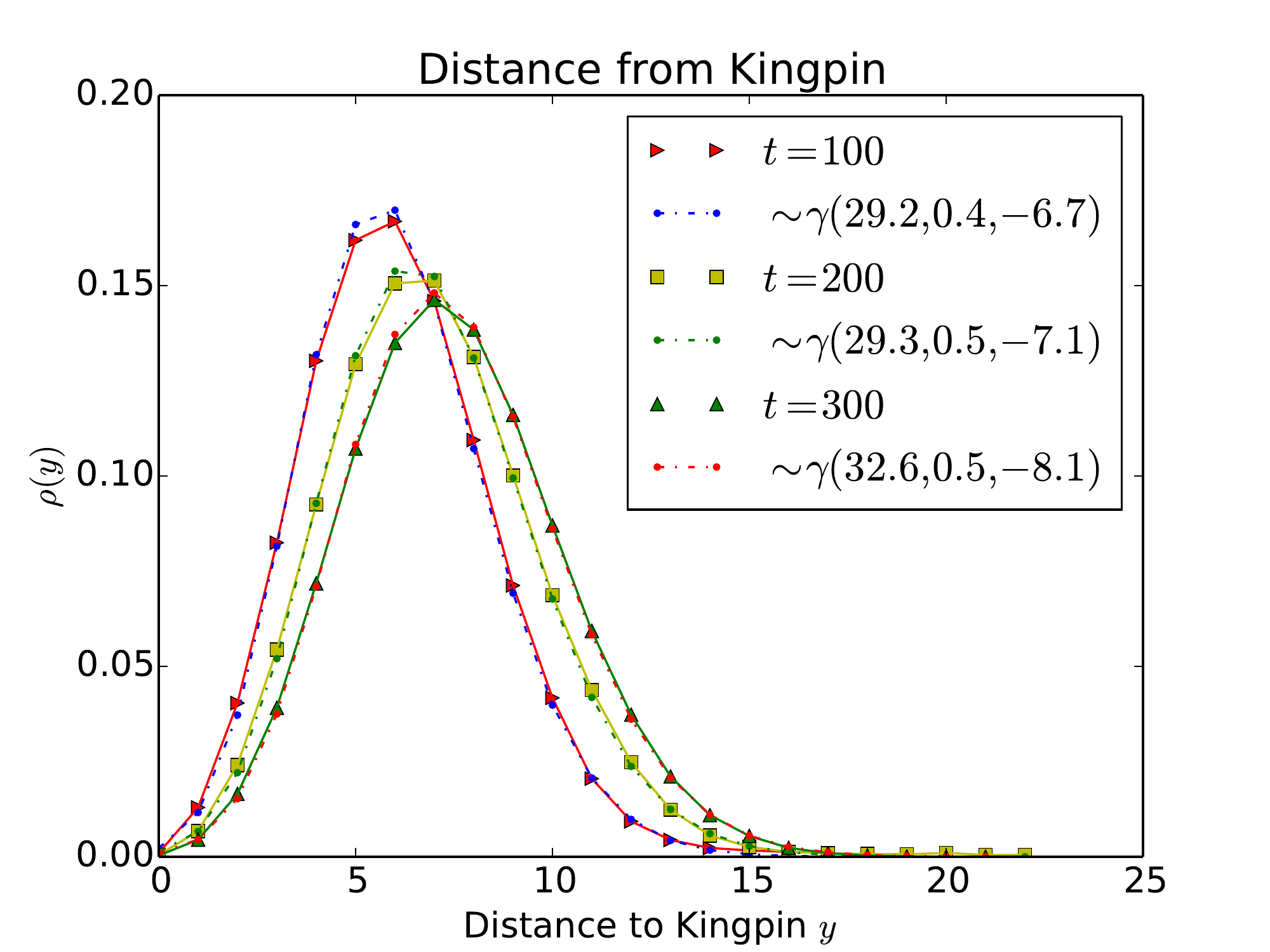}
\caption{
The distribution of criminal position relative to the kingpin after 100 runs.  The three curves represent the different times $t$ at which the recruitment was stopped.  Here, $t = 100, 200, 300$.  The recruitment index was set at $k=5$ and the initial configuration of the network was that of the single kingpin.  The probabilities were fit using the shifted gamma density $\rho_{\alpha, \beta, s}(y)$ found in Eq. \ref{density} and with parameters $\gamma (\alpha,\beta, s)$
specified in the legend. Similar shaped curves arise for larger values of $k$.}
\label{nd}
\end{figure}

\noindent
We now investigate how criminals are positioned relative to the kingpin, at the core of the network. We expect the distribution of criminals relative to the kingpin to become more uniform as the network grows in scale.  In Fig.\,\ref{nd} the recruitment process is stopped at a fixed time $t$, when we measure 
$\rho (y)$, the ratio of criminals at a distance $y$ from the kingpin with respect to the total number of nodes. Our measured $\rho(y)$ is then fitted to a shifted gamma probability density, the continuous analog of negative binomial distribution
\cite{chayes} and given by

\begin{align}
\rho_{\alpha, \beta, s}(y) = \frac{\beta^\alpha}{\Gamma(\alpha)}(y - s)^{\alpha-1}e^{ -\beta (y-s)}\label{density}
\end{align}
for $y > s$ and $\alpha > 1$. Our choice was motivated by the fact that this distribution is supported only on a portion of the horizontal 
axis. Using the fitted values of $\rho_{\alpha, \beta, s} (y)$ the probability that a criminal is a distance $y$ from the kingpin can be estimated as

\begin{eqnarray}
\int_s^y \rho_{\alpha, \beta, s}(y') \; dy'.
\end{eqnarray}

\noindent
From Fig.\,\ref{nd} we note that as $t$ increases, the average distance from the kingpin increases and that the distribution
of criminals becomes broader, as can be expected.  We also used different initial conditions, starting the recruitment process on 
given, already established networks and found that, at long times, the shifted gamma probability density remained a valid approximation
for  $\rho(y)$. 

\subsection{Street Criminals}
\begin{figure}\centering
\includegraphics[width=1\linewidth]{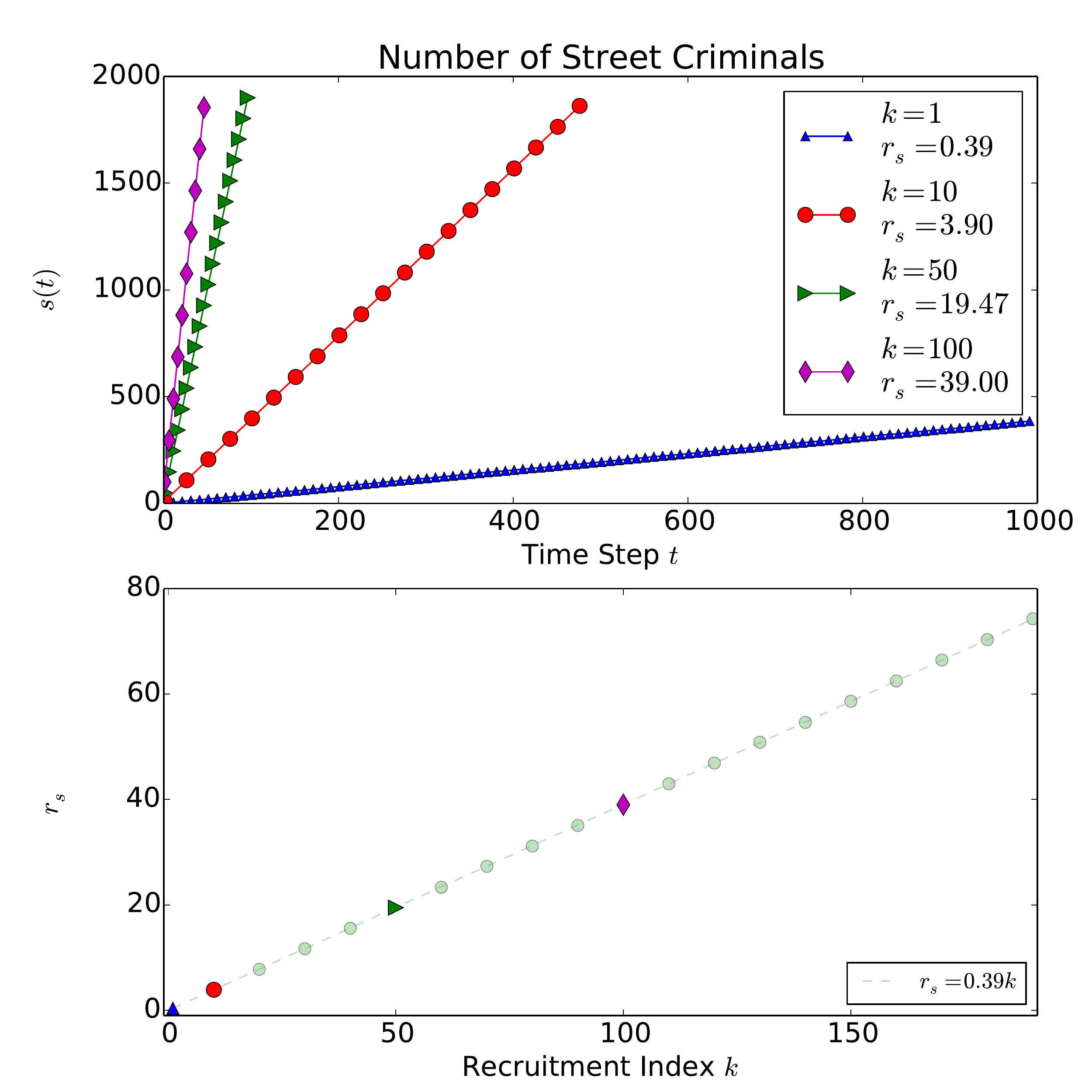}
\caption{(Top) Number of street criminals  $s(t)$ as a function of time for recruitment rates  $k = 10, 50, 100$.  In each case, the recruitment process was terminated when the total number of criminals exceeded $5 \times 10^3$ and averaged over 100 runs. We fit the data to $s(t) = r_s t + 1$ and expect $r_s \simeq P_{\infty}(d=0) k$ with the universal factor $P_{\infty}(d=0) = 0.39$. This scaling is confirmed by the fitted values of $r_s$ as can be seen from the values shown in the legend. (Bottom) The slope values $r_s$ as a function of $k$ with the $r_s$ values in the top display shown in dark symbols.  This data is then linearly fit as shown in the bottom legend, confirming our conjecture $r_s \simeq P_{\infty}(d=0) k $.}
\label{leaf}
\end{figure}

\noindent
In our model, street criminals are the nodes in ${\cal S}(t)$ without any underlings. We assume theirs is the only activity to be visible to law enforcement
making street criminals the most vulnerable to arrest.  At the same time, due to the choices made for $w(j;t)$, street criminals also have the highest probability of recruiting new members into the organization.  Since the network grows linearly in time, we expect the total number of street criminals $s(t)$ to increase accordingly, and that the proportion of street criminals with respect to the total number of nodes will remain fixed at the universal value
 $P_{\infty}(d=0) = 0.39$ as shown in Fig.\,\ref{DD}. In Fig.\,\ref{leaf} we plot $s(t)$ and fit the data to a linear form $s(t) = r_s t + 1$, where the unitary intercept is chosen since at $t=0$ the only criminal present is the kingpin, who does not have any underlings yet. We expect $r_s \simeq P_{\infty} (d=0) k $ as verified in the lower panel of Fig.\,\ref{leaf}. 
  
\begin{figure}[t]
\centering
\includegraphics[width=0.7\linewidth]{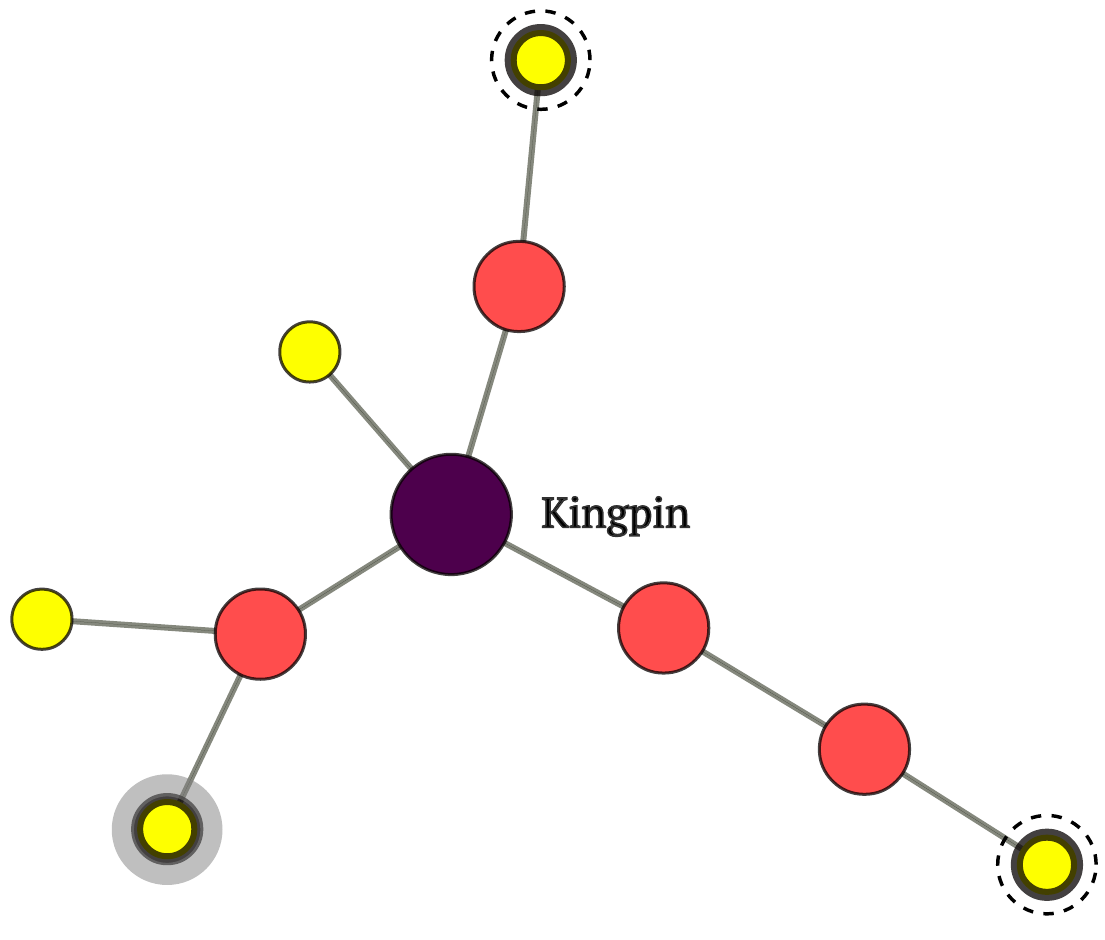}
\caption{Schematic of the addition of new nodes from time $t$ to time $t+1$. Here, $k=3$ and the new nodes are depicted in 
light yellow color. The addition of a street criminal to a more senior one, represented
by the node surrounded by a solid ring on the lower left hand side, will increase
 $s(t+1)$ by one unit with respect to $s(t)$. Vice-versa, the addition of a new node to an existing  street criminal, as shown by the nodes surrounded by the dashed circles,
will not change the number of street criminals, since for every new member of $\c S(t+1)$, one from $\c S(t)$ will be removed.  }
\label{harmseries}
\end{figure}

We can also write an iterative equation for $s(t+1)$ \cite{newmanbook, durrettbook}. At a given time $t$ the likelihood of adding a new street criminal to the network is given by the probability 
of adding a new node to a senior criminal, one that already has underlings. Conversely, the net number of street criminals will not change 
upon adding a new node to existing street criminals, since for every new street criminal added to the enumeration, there will be one that will be removed, having added a new underling. This is illustrated in Fig.\,\ref{harmseries}.
The weight associated to adding a new node to a senior criminal is given by 

\begin{eqnarray}
\label{senior}
\sum_{j \in \c [\c C(t)-\c S(t)]} w(j;t) =  \sum_{j \in \c  C(t)} w(j;t) - \frac {s(t)}{a}
\end{eqnarray}

\noindent
since the number of street criminals is $s(t)$ and their weight is given by $1/a$. The total probability of adding a criminal to a senior node
is thus given by Eq.\,\ref{senior}, normalized with respect to the total weight $\sum_{j \in \c C(t)} w(j;t)$. We can now write our iterative equation for
$s(t+1)$. The number of added new street criminals is given by the probability of adding a street criminal to a senior one as described above,
 multiplied by the total number of available new criminals at each time step, given by $k$. Using $a=1$ we find
  
 \begin{eqnarray}
 \label{scale0}
s(t+1) \simeq s(t) +  \frac{\sum_{j \in \c  C(t)} w(j;t)-s(t)}{ \sum_{j \in \c  C(t)} w(j;t) }  k. 
\end{eqnarray}

\begin{figure}[t]
 \centering
\includegraphics[width=1.0\linewidth]{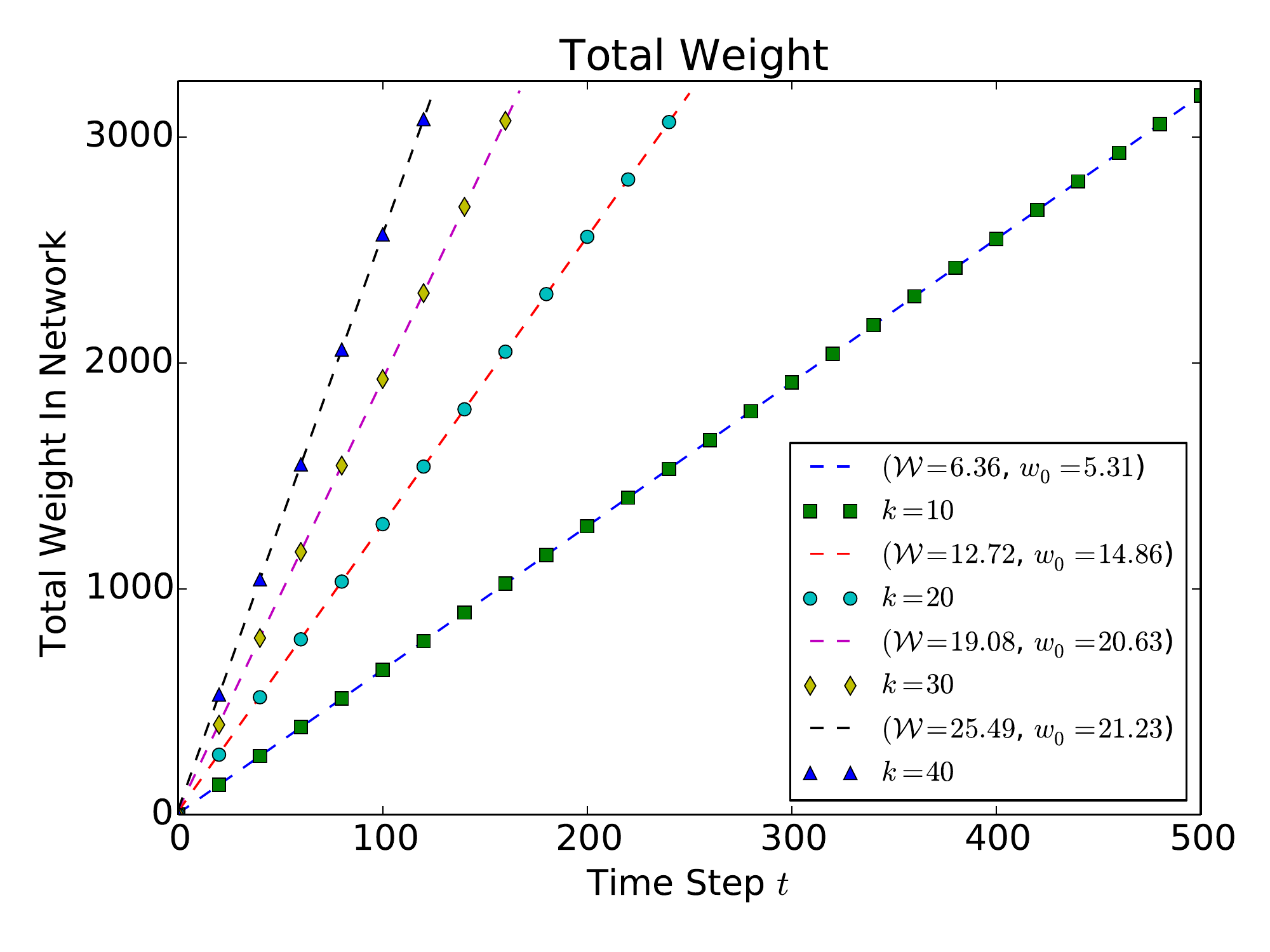}
\caption{The total weight of the network $\sum_{j \in \c C(t)} w(j;t)$ as a function of time averaged over 100 runs for $k=10,20,30,40$.
At the onset the network consisted of the kingpin alone; simulations were stopped when the total number of criminals exceeded $5 \times 10^3$. 
We fit the data to a linear form $\c W t + w_0$ and find that the extrapolated values of $\c W$ shown in the figure legend are in excellent agreement with the ones predicted from
Eq.\,\ref{scale2}, given by $\c W =6.39, 12.79, 19.18, 25.57$ for $k=10,20,30,40$, respectively.}
\label{totalweight}
\end{figure}

\noindent
We can use this relationship to heuristically determine the weight of the entire tree  
$\sum_{j \in \c  C(t)} w(j;t)$ at long times. The latter can be assumed to scale linearly, since the total number of
members of the network grows at the same rate, and at least a fraction of nodes $P_{\infty}(d=0)$ will be associated to 
a finite, unitary weight. We thus posit 
 $\sum_{j \in \c  C(t)} w(j;t) \simeq \c Wt$ as $t \to \infty$. We also use $s(t) \simeq  r_s t$  as $t \to \infty$
so the recursion relation Eq.\,\ref{scale0} at long times becomes

 \begin{eqnarray}
 \label{scale}
r_s  \simeq \frac{\c W -r_s }{ \c W  }  k , 
\end{eqnarray}
 
\noindent
yielding

\begin{eqnarray}
\label{scale2}
\c W  \simeq \frac{k r_s}{k - r_s} \simeq \frac{k P_{\infty}(d=0)}{1 - P_{\infty}(d=0)},
\end{eqnarray} 

\noindent
and where the last term is obtained via $r_s \simeq k P_{\infty}(d=0)$.
In Fig.\,\ref{totalweight} we plot the total weight of the network as a function of time for $k=10,20,30,40$ 
and find that it scales linearly, as we had assumed. We also find that the corresponding numerical fits 
yield good agreement with the estimates from Eq.\,\ref{scale2}.

\section{Police Pursuit}

\noindent
In this section, we introduce police agents to our model and describe the pursuit mechanisms they are engaged in on a network that is concurrently growing
in time.  The ultimate goal of law enforcement is to reach the kingpin, capture him or her, and dismantle the expanding criminal organization.  The ultimate goal of the criminal enterprise is to expand as much as possible. As discussed earlier, it is reasonable to assume that the global structure of the network is unknown to law enforcement agents who can begin their ``investigative" activities only at the bottom of the network, populated by street criminals.  Once a criminal network is formed, at every time step $t$  we choose a random street criminal in $\c S(t)$ as the initial search node for the officer.  This street criminal is now considered to be under investigation. The officer can decide whether to ``arrest" the criminal in question or move to one of its associates, chosen among the nodes that are linked to the current suspect.  In the next subsection we will illustrate three different ways of making this strategic decision. For now, we note that if the choice to arrest the current suspect is made, the latter is removed and this particular investigative round at time $t$ is complete. If the choice to migrate to a linked criminal is made, the pursuit continues: the officer is now faced with the same decision on whether to arrest or continue investigating.  We impose that a node that has already been visited by law enforcement cannot be visited again.  A sequence of investigative choices thus lead to a self-avoiding random walk on the criminal network at time $t$. The pursuit ends if an arrest is made or if the officer reaches either the kingpin or another street criminal.  

If the kingpin is reached, we consider the attempts of law enforcement to be successful: the criminal organization has been dismantled at time $t$ and the process is terminated. Vice-versa, if the investigative unit ends on another criminal in $\c S(t)$, we consider law enforcement intervention at this time to have failed: no arrests will be made, and the network stays unchanged. Finally, if law enforcement decides to perform an arrest on a given node, all of its underlings in the hierarchal structure will be removed as well. Note that since the node at which the arrest takes place can be a few links removed from the original starting point, an arrest does not necessarily imply that the first street criminal to be investigated will be removed as well.  The case of an arrest may be considered a partial success, since eliminating a few nodes on the hierarchical structure may make the kingpin more vulnerable in future time steps. After the pursuit phase at time $t$ is completed and assuming the kingpin has not been reached, the criminal network is grown according to the recruitment methods described in the previous section.  Pursuit and recruitment are then iterated at time $t+1$ and until the network reaches a given size $n^*$, with the network evolving dynamically in time.   Note that since at the onset of our simulation the only criminal present is the kingpin, we do not introduce the police pursuit at time $t=0$, since the kingpin would be arrested immediately and the criminal organization would not grow. Rather we consider an existing network, as shown in Fig.\,\ref{pe} for a full ternary tree of height three with forty criminals, as the initial configuration on which the pursuit is started.  
A full ternary network is one where all nodes, except street criminals, have exactly three underlings. A complete ternary network is one where all
nodes, except street criminals and the nodes immediately above them, have exactly three underlings. We use these full or complete networks
as initial conditions for an ``unthreatened" criminal organization, prior to police intervention, and assume that once law enforcement investigations have begun the network grows according to the recruitment rules described in the previous section. 

Our pursuit model could be adjusted to model different scenarios. For instance, a biased random walk might be more appropriate to describe the motion of law enforcement
if disruptors are privy to intelligence  regarding the hierarchical criminal structure, or they may wish to re-investigate criminals during their pursuit based on incoming
information.

\subsection{Pursuit Strategies}

\begin{figure*}\centering
\includegraphics[width = .8\linewidth]{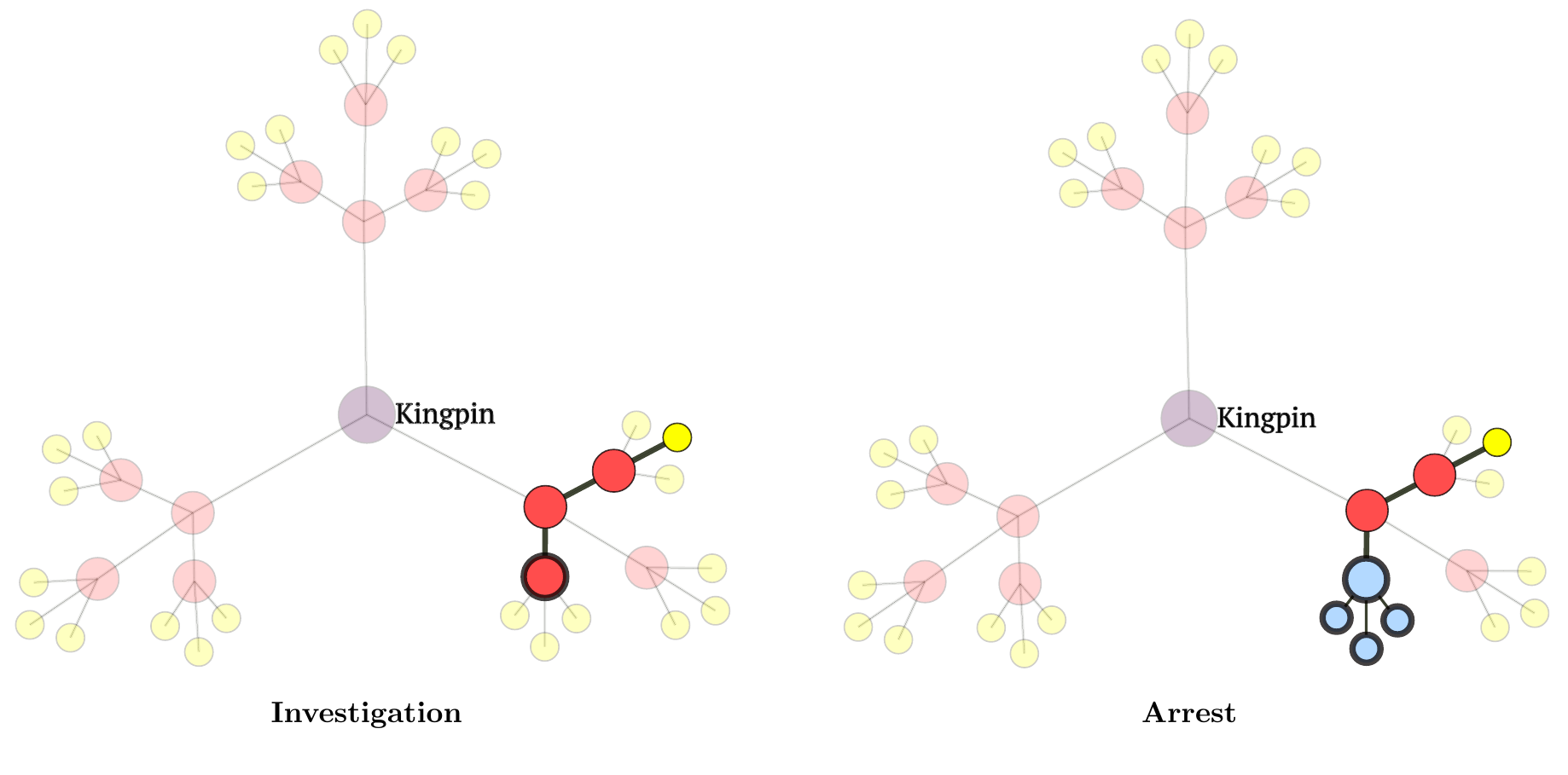}
\caption{The pursuit process as described in our text.  We start the network as a full ternary tree of height three with forty criminals, 
as shown by the light colored nodes.  
(Left) At time $t=1$ a law enforcement agent begins an ``investigation" from the highlighted yellow criminal and without having full knowledge of the network.
The investigative trail involves three other nodes, highlighted in red and linked by a self-avoiding random walk, marked by a solid line. The last node, surrounded by a dark ring, is the criminal that is arrested. (Right) Once a criminal is arrested and removed from the network all related underlings in the hierarchy are removed as well. In this case, all removed nodes are depicted in blue and have a darker boundary.  We note that not all criminals arrested were investigated and vice-versa.  }
\label{pe}
\end{figure*}

\noindent
We now discuss the possible strategies that law enforcement may employ when deciding whether to arrest a criminal or 
continue investigating other, neighboring nodes.  As already discussed, the police is assumed not to have full knowledge of the
entire organization but only of the criminals under investigation and their linked nodes. Because this is a dark network, 
the pursuit process may be unsuccessful with law enforcement
reaching street criminals that are low on the criminal hierarchy, representing a dead-end.

The first possible pursuit method we consider is the fixed investigation number strategy, where starting from a given street criminal, a disruptor will investigate $p$ successive nodes before making an arrest, assuming it has not encountered any new street criminals along its self-avoiding random walk.
This strategy will be denoted by $Q_{A}(p)$. The second method is the minimum out degree strategy, where law enforcement officials will keep investigating until a node of  at least out degree $q$ is reached, similarly assuming the self-avoiding random walk does not lead to other street criminals.  In this case, it is assumed that police agents are seeking to maximize the influence of the criminal to be arrested,
since the higher $q$ is, the more direct underlings the suspect will be affiliated to. We denote this strategy by $Q_{D}(q)$.
Finally, we consider the persistent investigative strategy where the pursuit is stopped
only upon reaching the kingpin or a dead-end street-criminal. This third and final strategy will be denoted by $Q_I$.  In Fig.\,\ref{pe}, we show a disruptor
 pursuing strategy $Q_A(q=3)$ on an initial full ternary tree of height three.  
 
The above pursuit strategies reflect certain real-world objectives of law enforcement.  $Q_A(p)$ for example, may be used by agencies that regularly make arrests, possibly  after investigation resources have been depleted, to show results to the general public in order to garner support  \cite{mexico2}. Signaling the presence of authorities via periodic arrests may also serve to deter the network's growth  \cite{broken_windows}.
$Q_D(q)$ may reflect the oft-pursed strategy of capturing high ranking criminals to optimally disrupt strong hierarchical operations \cite{wilson, oc_book}.
Lastly, $Q_I$ can represent law enforcement's choice to minimize violence. Since the arrest of a high ranking criminal could lead to deadly confrontation or to a network's violent reorganization, keeping all investigative operations covert until the kingpin is reached may generate the least violence \cite{cartels_, pablo1}.  Of these strategies, we expect  $Q_I$ to be the most expensive, as it either eradicates the network by kingpin capture or the network is left untouched. Since no intermediate arrests are performed, the network can effectively grow as if law enforcement were not present until the kingpin is reached.  The above strategies are related by 

\begin{figure*}\centering
\includegraphics[width=1\linewidth]{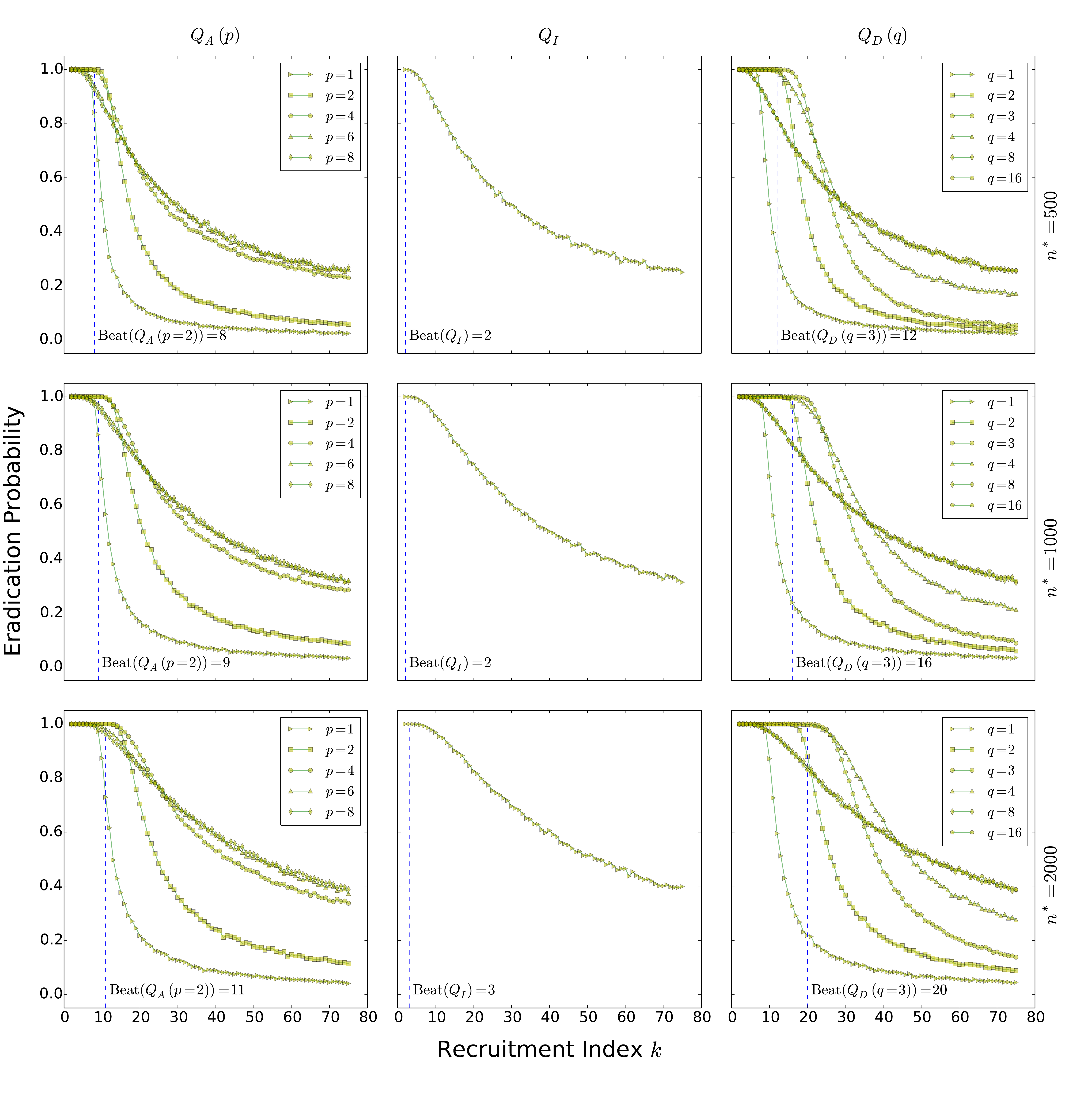}
\caption{The network eradication probability as a function of the recruitment index $k$, obtained by averaging over 10,000 simulations for different strategies. We consider a total population of $n^*=500, 1000, 2000$ individuals and halt our simulations when the criminal network reaches this size.
We specify the Beat($Q$)
of each strategy as the maximum value of the recruitment index $k$ for which the network is eradicated with probability one, over all simulations.
Note that $Q_I$ is the limiting strategy for $Q_A(p \to \infty)$ and $Q_D(q \to \infty)$. Our results reveal that the optimal strategy for
fast growing networks with large $k$ is to use investigative strategies with large values of $p,q$ 
while for slower growing networks with small values of $k$, moderate
values of $p,q$ yield higher probabilities of eradicating the network. Note that curves for $Q_A(p=1)$ and $Q_D(q=1)$ are the same and that the $Q_I$ curve is the limit for $Q_A(p \to \infty) = Q_D(q \to \infty)$.}
\label{beat}
\end{figure*}

\begin{eqnarray}
\lim_{p \to \infty} Q_A(p) = \lim_{q \to \infty} Q_D(q) = Q_I, \\
Q_A(1) = Q_D(1).
\label{lim2}
\end{eqnarray}

\noindent
Note that as the parameters $p$ and $q$ increase, the chosen strategies become more and more covert and demanding, 
involving more investigations and aiming at higher level arrests, so that 
$Q_A$ and $Q_D$ approach $Q_I$.  Also, note that under strategy $Q_A(1)$ law enforcement will remove the node directly above the 
street criminal selected as the initial investigative point. This is the same result that would arise from strategy $Q_D(1)$, since all nodes
above street criminals have at least one linked criminal, \textit{i.e.} the street criminal itself.

\begin{figure*}
\includegraphics[width=1\linewidth]{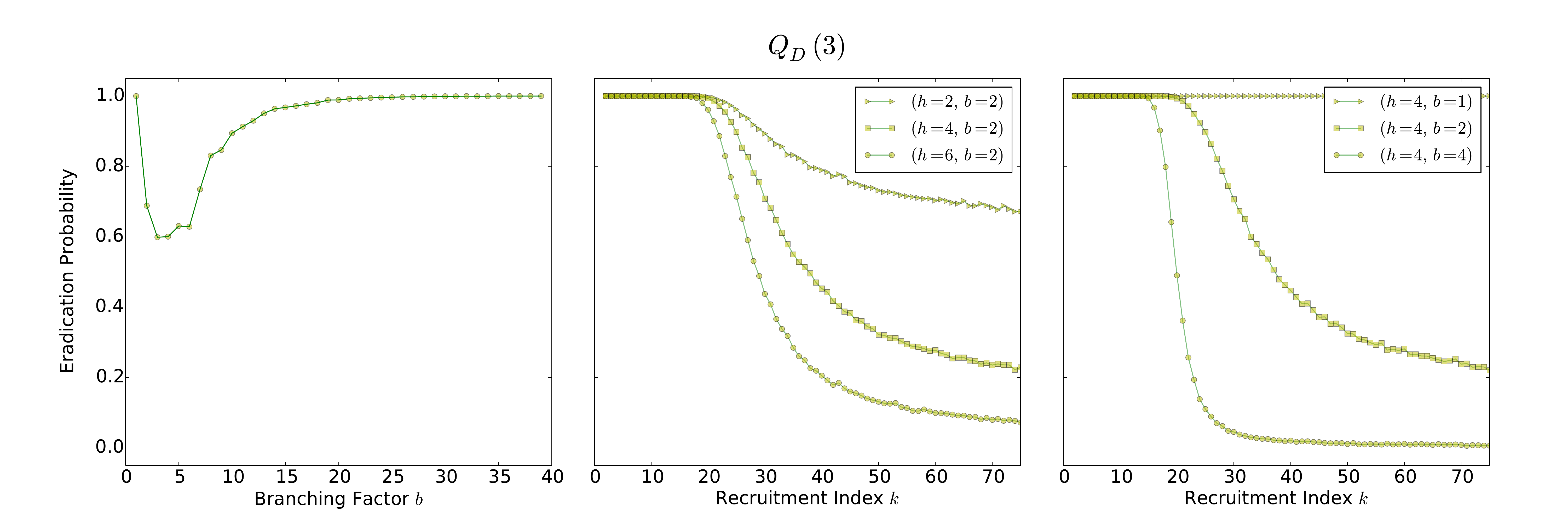}
\caption{ Varying initial conditions prior to law enforcement intervention.
(Left) Eradication probabilities on a complete initial tree with $b$ branches and forty criminals for $k=30$ and law enforcement
strategy $Q_D(q=3)$. On a linear chain ($b=1$) the eradication probability is one, since no dead-ends will be reached. Increasing
$b$ allows for more dead-ends to be encountered so that the eradication probability decreases until a threshold value of $b$
 when the height of the network becomes small enough to allow for easier access to the kingpin.  Here $b=3$.
(Right) Eradication probabilities using full initial trees with $b$ branches and height $h$ as initial conditions and using
strategy $Q_D(q=3)$.  The initial number of criminals is $(b^{h+1} -1) /(b-1)$.  For all values of
$k$ eradication is higher for lower values of $b,h$ indicating that best results will be obtained with law enforcement intervening
on initially contained networks, as can be expected. 
Qualitatively similar results arise for other strategies $Q_D(q)$ and $Q_A(p)$.}
\label{ic}
\end{figure*}

We can evaluate each strategy's performance for several recruitment indices $k$ using numerical simulations. Runs were continued until either the kingpin was arrested, eradicating the criminal network,  or the kingpin was not arrested and the network exceeded a total, given population size $n^*$. The latter scenario represents the case of a vast organization permeating all society. In Fig.\,\ref{beat} we plot the network eradication probability for $Q_A(p), Q_D(q) $ and $Q_I$ as a function of the recruitment index $k$,  for various choices of $p,q$ and for various thresholds of criminal population $n^*$.
In all cases, as can be expected, the probability of capturing the kingpin decreases with  $k$, as the rate of adding new criminals becomes faster than any disruption attempts by law enforcement. We also define $\t{Beat}(Q)$, the ``beat number" of strategy $Q$, as the maximum recruitment index $k$ of the network for which law enforcement will reach the kingpin with unit probability across the simulations performed.
The eradication probabilities and $\t{Beat}(Q)$ depend on the total size of the network $n^*$ as can be seen from
Fig.\,\ref{beat}. $\t{Beat}(Q)$ increases with $n^*$ especially for small and intermediate values of
$k$. Here, growth occurs at a relatively slow rate and although increasing $n^*$ allows for more criminals to join the organization, 
there will also be a relatively large number of police pursuits during the slow dynamics, allowing for greater
eradication probabilities. Given a fixed value of $n^*$, we can compare different strategies and their results. 
From the left hand panels of Fig.\,\ref{beat} that illustrate results for the fixed investigation number strategy $Q_A(p)$
we note that for large $k$ values the eradication probability increases
with the number of investigations $p$. In this case the network is rapidly expanding and allowing more investigations
before an arrest is made also allows for the possibility of arresting senior, highly nested criminals 
with many underlings, greatly undermining network structure and size.  
For small values of $k$ few criminals are added at each time step and each node will have few underlings. Small values of $p$ restrict law enforcement to activity close to street level,  where nodes have low hierarchical value and allow for modest but progressive node removal. 
Increasing $p$, when $p$ is small, is beneficial as can be seen by comparing the $p=1,2$ curves. However, when $k$ is small, 
increasing $p$ to larger values may not be the best strategy: since each node has few underlings, higher values of $p$ increase the possibility that the self-avoiding random walk performed by law enforcement reaches a dead-end, effectively leaving the network untouched. 
This is the case for example for $Q_A(p=6)$ and $Q_A(p=8)$ for which the eradication probability is smaller
than for $Q_A(p=1)$ and $Q_A(p=2)$ for small $k$ values.
Similar trends arise for the minimum out degree strategy
$Q_D(q)$ as shown in the right hand panels of Fig.\,\ref{beat}. Here, for large values of $k$ the best strategy is 
to set a target of relatively large $q$ before performing an arrest,  
while for lower values of $k$ moderate $q$ values are preferred, due the possibility
of reaching dead-ends during the pursuit. The intermediate panels in 
Fig.\,\ref{beat} show results stemming from the persistent investigative strategy $Q_I$, which corresponds to $Q_A(p \to \infty)$
and $Q_D(q \to \infty)$. Comparing the respective panels of Fig.\,\ref{beat} shows that the optimal approach from the perspective of law enforcement, whether engaged in the fixed investigation number pursuit $Q_A(p)$ or in the minimum out degree strategy $Q_D(q)$, is to adjust arrest criteria to an optimal $p^*$ or $q^*$ 
depending on the recruitment rate $k$, if this variable is known or estimates are available.

The results discussed so far depend on initial conditions, which can be chosen to be
any full or complete tree with $b$ branches and height $h$.  To analyze different initial conditions, we can vary 
the values of $b,h$ that describe the ``unthreatened", police-free tree. Note that full trees
have a total number of $(b^{h+1} - 1)/(b-1)$ criminals; the case of a full ternary tree of height three discussed above
corresponds to $b=h=3$ with forty initial criminals. For a fixed number of criminals, the larger $b$, the smaller $h$ and the less hierarchical
the initial tree is. The choice of $b=1$ is the limiting case of
a linear chain: here, strategy $Q_D(q>1)$ will lead to a unitary eradication probability 
since all nodes will have out degree one and at the onset of the pursuit phase law enforcement will proceed until
the kingpin is reached, regardless of the number of total criminals and of $h$.  The eradication probability decreases
on a binary initial tree with $b=2$, since now dead-ends may be encountered.  Further increasing $b$ leads to diminishing
eradication probabilities due to the possibility of more unsuccessful pursuits. Once a sufficiently large value for $b$ is reached
however,  $h$ will be small enough, that reaching the kingpin may become more feasible. These two opposing trends are expected to lead to a
minimum in the eradication probability.  We find the critical value of $b$ at the threshold between the two trends to depend on the initial number of criminals. In the left hand panel of Fig.\,\ref{ic}  for example we show the eradication probability for $Q_D(q=3)$ on an initial network of forty criminals with varying $b$ and with a fixed value of $k=30$. 
The complete trees we created were as regular as possible,
with all second-to-last nodes sharing the same number of underlings as possible.
The choice $b=3$ is the full ternary tree case we have analyzed in detail 
above. Note the minimum for intermediate values of $b$. Similar trends arise also for $Q_A(p)$ although the decrease in the eradication probability is less pronounced for small $b$. 
In the middle and right hand panels of Fig.\,\ref{ic} we consider initial networks made of full trees with fixed $b$ and varying $h$ and vice-versa.
We expect increases in $b$ or $h$ to lead to lower eradication probabilities since the number of initial criminals will be larger.
In the middle panel of Fig.\,\ref{ic} we show an initial network of height $h=4$ with branching $b=1$, representing a chain of criminals, with $b=2$,
representing a binary tree, and with $b=4$ representing a quaternary one. As expected, the eradication probability
decreases with $k$ for all $b$,  and with $b$ for fixed $k$. On the right hand panel of Fig.\,\ref{ic}
we find similar results for an initial full binary tree with $b=2$ and with heights $h=2,4,6$.  The eradication probability 
decreases with $k$ for all $h$, and with $h$ for fixed $k$ as well.
 
\begin{figure*}[t]
\centering
\includegraphics[width=\linewidth]{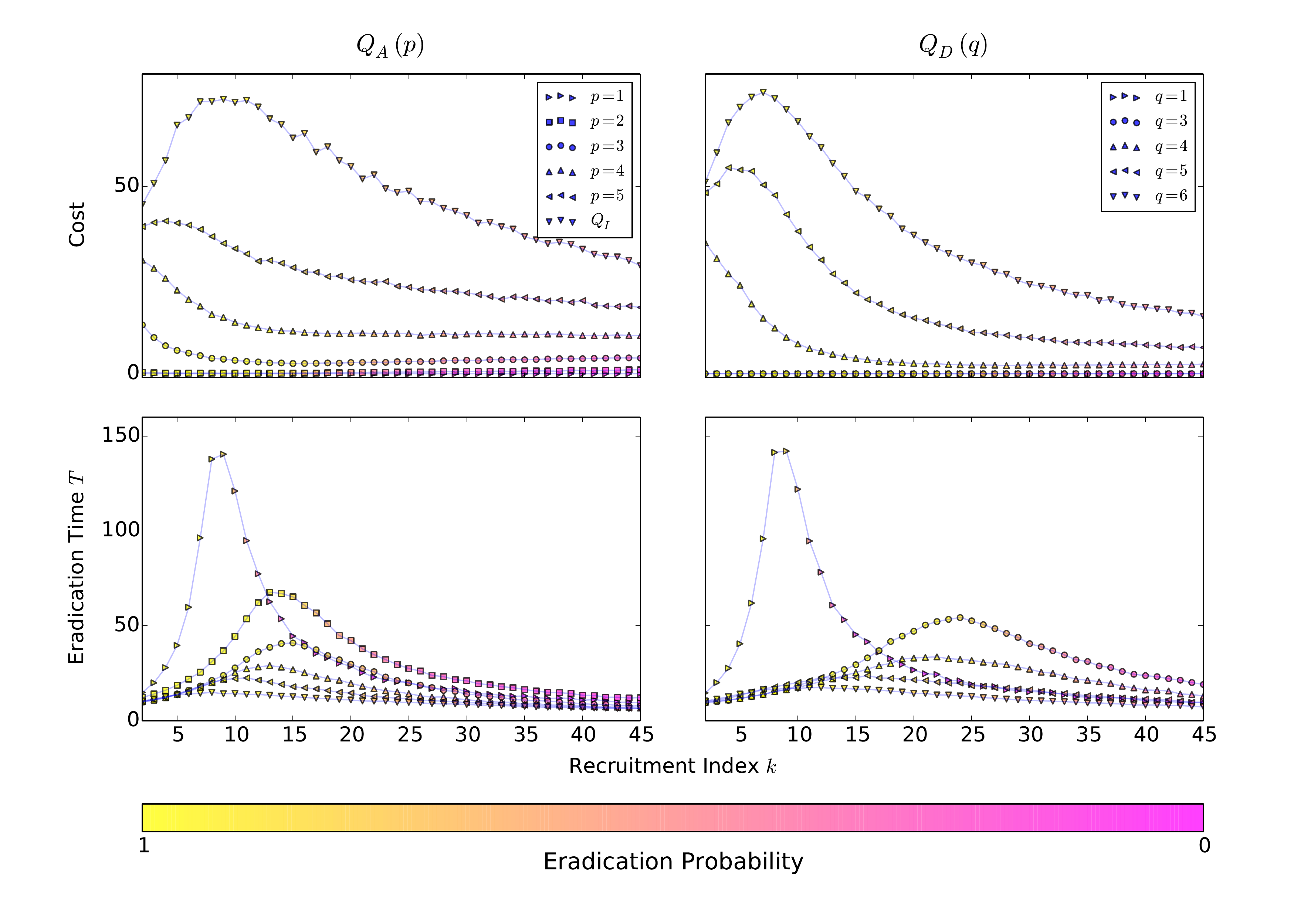} 
\caption{(Top) Costs incurred by law enforcement conditioned on kingpin capture as a function of the recruitment index $k$.  Shades in the  data points represent the probabilities of network eradication. We considered 10000 simulations and allowed the network to grow to $n^*=1000$. Note the emergence in maxima for all curves, due to the conditional nature of cost evaluations. As $k$ increases, the network grows more rapidly and so does the number of investigations necessary for eradication. Upon reaching a threshold in $k$, eradication becomes less likely especially in the latter growth stages so that for large enough $k$ either the kingpin is captured in the early stages of network growth, with little costs, or it will never will be. The requirement for quick capture with growing $k$ for large $k$ is associated with decreases in the cost function, leading to the maxima in $k$. (Bottom) The mean eradication time as a function of $k$ for various strategies. Similarly as for the above panel, the conditional nature of the process is manifest from the emergence of maxima in all curves. Note that curves for $Q_A(p=1)$ and $Q_D(q=1)$ are the same and that the $Q_I$ curve is the limit for $Q_A(p \to \infty)$.}
\label{cost}
\end{figure*}

 \subsection{Strategic Costs and Time to Eradication}

\noindent
In the previous subsection, we implicitly assumed that all pursuit strategies could be conducted with any value of $p,q$ indistinctly. Here we quantify the efficiency of various investigative methods by associating a cost measure to each of them. We assume that investigating criminals requires the expenditure of societal resources, while arresting criminals can be considered a gain, since an arrest will weaken the criminal network and the prospect of future crime will lessen. We thus evaluate cost as the number of investigations performed by law enforcement within a given pursuit phase minus the number of criminals removed within the same pursuit, if this difference is positive. Otherwise, if the number of investigations is lower than the number of arrests, we 
let costs be zero. For example, in the left hand panel of Fig.\,\ref{pe} four nodes are investigated, the street criminal (depicted in yellow), and three senior members (depicted in green). During the arrest phase, in the right hand panel of Fig.\,\ref{pe}, four criminals are eliminated (depicted in red). At this iteration, the costs associated with investigating four criminals are balanced by the benefits associated with removing four criminals, so we tally costs as zero. Total costs are then calculated as the cumulative cost required to reach the kingpin 
throughout the duration of the simulation, assuming the network is eradicated. If the network is not eradicated, we can still record the cumulative costs until the entire population $n^*$ has joined the criminal organization. However, we do not discuss cases where the kingpin is not reached, since here costs will be proportional to the number of rounds until the $n^*$ criminals are incorporated in the network.  Note that dead-ends in this context are very expensive, since there are no net gains in a pursuit that leads to no arrests.
In the upper panel of Fig.\,\ref{cost} we plot the total cost incurred by authorities conditioned on the network being dismantled
for a total population size $n^*=1000$ and for various pursuit methods $Q_A(p)$ and $Q_D(q)$ starting on an initial full ternary tree.
We also depict the probability of kingpin capture and criminal network eradication as shades in the data points.  Costs are identically zero for $Q_A(p =1)$ and $Q_D(q= 1)$ since in these cases there will be two investigations and two criminals removed at every time step. Similarly, costs stay low for low values of $p,q$ across both strategies. For example for both $Q_A(p =2)$ and $Q_D(q= 3)$ total expenditures are very small, indicating that the number of investigations is at most comparable to the number of criminals removed throughout the entire network evolution.

For intermediate values of $p,q$  we note that curves in the upper panels of Fig.\,\ref{cost} 
are monotonically decreasing, as can be seen for $Q_A(p =3), Q_A(p=4)$ and $Q_D(q=4)$.  
To understand this behavior, we note that several distinct trends arise upon increasing $k$ for small values of $k$. On one hand, the likelihood 
of reaching costly dead-ends increases. Here, larger values of $k$ represent faster network growth, with more nodes
being added at every time step. Each criminal is thus linked to a greater number of underlings and the probability of moving higher up in the hierarchy decreases. In this case, it is very likely for law enforcement to eventually reach a dead-end, and for costs to increase. On the other hand, if dead-ends are not reached and arrests are performed, larger $k$ values imply that more underlings will be eliminated per pursuit, leading to decreasing costs.  
Of these two trends, for intermediate values of $p,q$, the most important is the greater elimination of criminals with increasing $k$. Here, 
pursuits are short-lived so that the self-avoiding random walks performed by law enforcement have a slightly lower chance of incurring in dead-ends, compared
to higher values of $p,q$. As a result, the number of criminals arrested at each time step is contained, but almost always will criminals be arrested,
leading to the monotonically decreasing curves in the upper panels of Fig.\,\ref{cost}.
Conversely, as $p,q$ increase further, pursuits are longer with dead-ends becoming more likely and more costly, failing to limit network growth.  Here, increasing $k$ for small $k$ leads to higher costs, as can be seen from the  $Q_A(p =5)$ and $Q_D(p=5), Q_D(p=6)$ curves. Note that the same trend arises for $Q_I$, which is the limiting behavior for $Q_A(p \to \infty)$ and $Q_D(p \to \infty)$. As $k$ increases even further, although the likelihood of reaching dead-ends increases, the gains in eliminating more criminals prevails and all curves show decreasing costs with increasing $k$ leading
to a maximum for intermediate $p,q$. 

These behaviors are valid for small and intermediate values of $k$, when the likelihood of eradication is almost unitary. 
Beyond a certain threshold in $k$ however, the eradication probability decreases for all choices of $p,q$ and 
the probability that the entire society is taken over by organized crime increases. In this case, either the network is eradicated at its early stages of growth, or it will never be.  Compounded with the above considerations, the decrease in the cost curves for all values of $p,q$ for large $k$ are indicative of the conditional nature of the process: as $k$ increases beyond a certain threshold, it becomes less and less likely to  be able capture the kingpin and the process must occur more and more swiftly with less investigations, so that costs decrease as a function of $k$ regardless of $p,q$. 
In the lower panel of Fig.\,\ref{cost} we plot the time of first eradication of the network as a function of recruitment index $k$ and find trends
that support these considerations.  The time of first eradication is always non-monotonic: it increases in $k$ before decreasing again, with 
the non-monotonicity stemming from the conditional manner in which eradication times are evaluated. 
For small $k$, increasing $k$ requires more time steps in capturing the kingpin but beyond a certain threshold, eradication must be quick or it will never
occur at all, leading to decreasing first eradication times. Indeed, peaks in the eradication time curves correlate to drops in the eradication probability, as can be seen from the shaded colors in the lower panels of  Fig.\,\ref{cost}.
Note that for small to intermediate $k$ values  increasing eradication times may be coupled with lower costs, indicating that while more time steps are required to reach the kingpin,  criminals are being eliminated from the network in a more efficient way.

It is important to note that these results are highly dependent on the network configuration on which the pursuit was initiated: different initial conditions will yield different eradication probabilities, costs and eradication times. We simulated different initial tree configurations and
found that given an initial number of criminals, here set at forty, more hierarchical structures (complete linear or binary trees, with lower $b$ values) 
allow for better results in terms of maximizing eradication probabilities and minimizing costs and eradication times as discussed in the previous subsection.
In general, initial networks a lower number of initial criminals, with either lower $b$ or $h$ allow for shorter eradication times and lower costs, as can be expected.

\subsection{Best strategies}

\noindent
From the above results we can try to identify a best strategy for network eradication. As discussed above,
eradication probabilities decrease with $k$. Law enforcement cannot influence $k$ values, as this is an intrinsic feature of the
criminal organization and may depend, for example, on kingpin charisma or on rewards offered by the network
to its members. Indeed, on dark networks, law enforcement may only have best guesses for $k$.
We thus assume that law enforcement agencies may only select which strategy to use given 
a preset value of $p=q=q^*$ that is not exceedingly large since for $q^*\to \infty$  all strategies are the same.  
Fig.\,\ref{beat} shows that for very small and very large values of $k$, 
given a full ternary tree as initial condition, eradication probabilities do not change significantly across
strategies. However, for intermediate values of $k$, the minimum out degree strategy $Q_D(q^*)$ is associated with slightly
larger eradication probabilities compared to the fixed investigation number strategy $Q_A(q^*)$. 
This can be seen for example by comparing curves for $Q_A(p=3)$ and $Q_D(q=3)$ and in particular by noting that Beat($Q_D(p=3)$) $>$ Beat($Q_A(q=3)$) for all values of $n^*$. From this perspective, given that typical $k$ values are not known to law enforcement, it is optimal to utilize strategy $Q_D(q^*)$ for a given value of $q^*$. The choice of what $q^*$ to select, if there is any information known on the rate of growth of the criminal organization is to use lower values of $q^*$ for slowly growing organizations, with lower values of $k$, and larger $q^*$  in the opposite case. If one is interested in lowering costs, for example when the criminal organization is not engaged in activities that are deemed to be especially dangerous for the community, the best intervention method is to use the less sophisticated investigative methods associates with lower values of $q^*$, since these are associated with lower costs, albeit to longer eradication times as well.  Costs are lower for $Q_D(q^*)$ than for $Q_A(q^*)$ as can be seen from Fig.\,\ref{cost}. Similarly, if one is interested in lowering eradication times, the minimum out degree strategy $Q_D(q^*)$ always yields better results than the fixed investigation number strategy $Q_A(q^*)$.

 \section{Conclusion}
 
 \noindent
The simple network model presented in this paper provides insight on the formation and active disruption of growing, dark criminal networks.  
We focused on organized, hierarchical drug syndicates, such as the Medell\'{i}n and Sinaloa drug cartels \cite{mexico, cartels_, cartels_2, oc_book, pablo1, sinaloa1, sinaloa2, escape}, some features of which were used to inform our modeling choices.  Other criminal associations that are hierarchically organized and that our model could be adapted to study
include the American and Sicilian Cosa Nostra mafia networks \cite{mob_code, mastrobuoni1, mastrobuoni2} and the Hells Angels biker gang  \cite{morselli}.

The recruitment mechanism we used is a variation on standard preferential attachment models, though there are some important distinctions. 
For example, the resulting out degree distribution we find is not heavy tailed \cite{heavy1, heavy2} and appears to be independent of the recruitment index $k$.  We also found the distribution of criminal position relative to the kingpin to be well approximated by a shifted gamma distribution with parameters depending on the initial configuration, on the recruitment index $k$, and on the maximum network size $n^*$. 
Our model yielded a linear relationship between the number of street criminals and the recruitment rate $k$, for which we provided a heuristic justification. A more rigorous framework could be useful 
to determine the relationship between criminals of higher degree than street criminals and $k$. This is a more difficult task than what presented in this work for street criminals, since the distribution would depend on the exact topography of the network.

Law enforcement pursuit and arrest were modeled as a dark network disruption problem.  We introduced and analyzed the efficacy of three investigative strategies that could be used:  $Q_D(p)$ with a preset number of investigations $p$ before an arrest is carried out, $Q_A(q)$ by seeking out a criminal with at least $q$ connections, and $Q_I$ by reaching the kingpin.  For a preset value of $p=q=q^*$ we can heuristically determine the most effective strategy on a dark network to be $Q_D(q^*)$, when the pursuit ends upon reaching criminals with at least $q^*$ connections and for moderate values of $q^*$. Indeed, strategy $Q_D(q^*)$ yields comparable or larger eradication probabilities than $Q_A(q^*)$. The optimal value of the arrest parameter $q^*$ will depend on the recruitment rate $k$ and on the overall population size $n^*$. For example, from Fig.\,\ref{beat} it appears that $Q_D(8)$ is more effective than $Q_D(3)$ only when $k \gtrsim 40$ for all values of $n^*$ with the opposite being true for smaller values of $k$.  Also in terms of minimizing costs and decreasing eradication times strategy $Q_D(q^*)$ is more efficient than strategy $Q_A(q^*)$ as can be seen from Fig.\,\ref{cost}. This result is consistent with previous models of fixed networks where the optimal disruptor strategy includes seeking the nodes of highest detectable degree \cite{mcbride1, mcbride2, dark_topo}.

We also find that as $k$ increases and the network grows at a faster rate, the eradication probability decreases and either the kingpin is captured
in the early stages of network growth or it will never be. This result provides a mathematical foundation to the common-sense notion of beneficial quick intervention and is in agreement with previous results on attempts to dismantle operational criminal networks. For example, in studies of drug trafficking networks in the Netherlands every disruption strategy proposed was largely ineffective except when performed during the nascent phase of the criminal organization \cite{duijn} .  While the disruption and recruitment processes analyzed in the latter work are different from ours, we can draw a similar conclusion on the importance of ``proactively" attacking organized crime networks before they become too entrenched within society and eradication proves more and more elusive. 

In this work we have only modeled a professional network of criminals. However, a social network may be more useful to law enforcement officials \cite{morsellibook}, with the possible inclusion of geo-spatial constraints.  Our preference mechanism can be generalized to directed networks using the length of the shortest path to the kingpin as input for node attachment when a tree structure is not present.  Moreover, adjusting the rate of criminal recruitment  might be more realistic for organized crime networks with external or economic pressures.  The disruption strategies presented here could also be made adaptive so that at some time threshold, strategy $Q_D(q)$ could be abandoned in favor of strategy $Q_A(p)$.  All of these could make this model more realistic and easier to validate using criminal network data.

Finally, we performed simulations with forty initial criminals varying the initial configurations prior to police intervention.  We found low branching, such as linear or binary trees, lead to higher eradication probabilities once law enforcement pursuits are introduced. Similarly extremely large initial branching lead to higher eradication probabilities. These results indicate that the simplest networks to eradicate are those with initially low $b$ where once investigations begin, at the early stages of the pursuit process, almost certainly the kingpin will be reached, or those that grow in an almost non-centralized manner where the many initial street criminals provided by very high $b$  allow for easy access to the kingpin who is just one or two levels removed from street activity. Extremely prudent (low $b$) or ambitious (high $b$) kingpins are thus the most vulnerable. On the contrary, the most robust initial networks are those that grow enough levels and with enough criminal members  prior to police intervention to effectively shield the kingpin from arrest as soon as the pursuit process is initiated, thus allowing for a vigorous successive growth.  

\section{Code}
\noindent  
We used NetworkX, NumPy, and SciPy for all the simulations.  d3.js was used for the network diagrams.  The code for the preferential attachment tree, the dynamic game, and the simulations can be found at \url{https://github.com/cmarshak/GameOfPablos}.\\
 
\section{Acknowledgements}
\noindent
We acknowledge fruitful discussions with James von Brecht,  Ryan Compton and Greg Klar.  This work was supported by ARO MURI grant W911NF-11-1-0332, AFOSR MURI grant FA9550-10-1-0569, NSF grant DMS-0968309, ONR grant N000141210838.

\FloatBarrier
\bibliography{bibliography}
\bibliographystyle{apsrev}

\end{document}